\documentclass{article}
\usepackage{spconf,graphicx}
\usepackage{amsmath,amsfonts,amssymb,amsthm, bm}
\usepackage{color}
\usepackage{float}
\usepackage{etoolbox}
\usepackage{multirow}
\usepackage{enumitem}
\usepackage{lipsum}
\usepackage{cite}
\usepackage{url}
\usepackage{xcolor}

\makeatletter
\def\@xfootnote[#1]{%
  \protected@xdef\@thefnmark{#1}%
  \@footnotemark\@footnotetext}
\makeatother

\newcommand{\paul}[1]{{\color{black}#1}}

\newcommand{\newadd}[1]{{\color{black}#1}}

\title{AUTOMATIC DJ TRANSITIONS WITH DIFFERENTIABLE AUDIO EFFECTS AND GENERATIVE ADVERSARIAL NETWORKS}

\name{\begin{tabular}{c}Bo-Yu Chen$^{1\ast}$\thanks{$\ast$ Work done during  an internship at Sony Group Corporation.}, Wei-Han Hsu$^{1}$,  Wei-Hsiang Liao$^{2}$, Marco A. Mart\'{i}nez Ram\'{i}rez$^{2}$, \\ \emph{Yuki Mitsufuji$^{2}$, and Yi-Hsuan Yang$^{1}$}\end{tabular}}

\address{$^{1}$ Research Center for IT Innovation, Academia Sinica, Taiwan \\
$^{2}$ Sony Group Corporation, Japan}

\begin{document}
\ninept

\maketitle

\begin{abstract}
    A central task of a Disc Jockey (DJ) 
    is to create a mixset of music 
    with seamless transitions between adjacent tracks.
    In this paper, we explore a data-driven approach that uses a generative adversarial network to create the song transition by learning from real-world DJ mixes. 
    The generator 
    uses two differentiable digital signal processing components, an equalizer (EQ) and a fader, 
    to mix two tracks selected by a data generation pipeline.
    The generator has to set the parameters of the EQs and fader in such a way that the resulting mix resembles real mixes created by human DJ, as judged by the discriminator counterpart.
    Result of a  listening test shows that the model can achieve competitive results compared with a number of baselines. 
\end{abstract}

\section{Introduction}
\label{sec:intro}

Recent years witnessed growing interest in building automatic DJ systems. 
Research has been done to 
not only computationally analyze existing mixes made by human DJs \cite{ reverse_dj,computational_dj,reverse_dj_transition}, 
but also develop automatic models that mimic how DJs create a mix \cite{djnet,drum&bass,spotify_transition,highlight_dj,reinforcement_dj}. 
We are in particular interested in the automation of DJ transition making, for it involves expert knowledge of DJ equipment and music mixing and represents one of the most important factors for shaping DJ styles.

\begin{figure}[t]
\centering
    \includegraphics[width=1\linewidth]{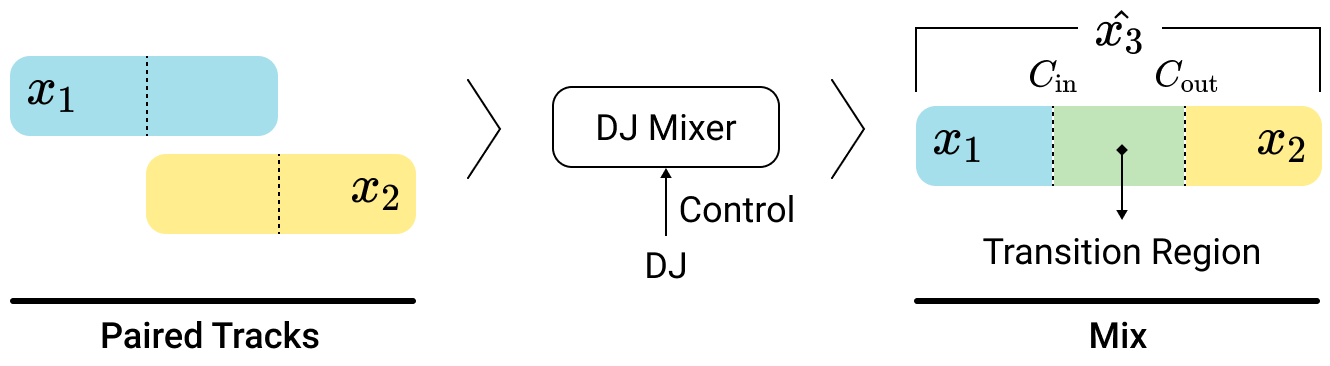}
    \caption{\paul{Illustration of the DJ transition making process.}}
\label{fig:transition}
\end{figure}

As depicted in Figure \ref{fig:transition}, the DJ transition making process is composed of at least two parts. Given a pair of audio files $x_1$ and $x_2$, \emph{cue point selection}  decides the ``cue out'' time ($C_\text{out}$) where the first track becomes inaudible, and the ``cue in'' time ($C_\text{in}$) where the second track becomes audible. Then, \emph{mixer controlling} 
applies audio effects such as EQ to the portion of $x_1$ and $x_2$ where they overlap (i.e., the \emph{transition region}), so that the
resulting mix $\hat{x}_3$ contains a smooth
transition of the tracks.
While cue point selection has been more often studied in the literature \cite{spotify_transition,cue_select}, automatic mixer controlling for DJ transition generation remains much unexplored.
Existing methods are mostly based on hand-crafted rules  \cite{drum&bass,highlight_dj}, which 
may not work well for all musical materials. 
We aim to automate this process by exploring for the first time a data-driven approach to transition generation, learning to DJing directly from real-world data. 

\paul{
To achieve this, a supervised approach would require the availability of the \emph{original} song pairs ($x_1, x_2$) as model input and the corresponding DJ-made mix $\hat{x}_3$ as the target output. Preparing such a training set is time-consuming for we need to collect the original songs involved in mixes, detect the cue points, and properly segment the original songs. To get around this, we resort to a generative adversarial network (GAN) approach \cite{gan} that requires only a collection of real-world mixes (as real data) and a separate collection of song pairs (called ``paired tracks in Figure \ref{fig:transition}) that do not need to correspond to the real-world mixes (as input to our generator). 
We use the principle of GAN to learn how to mix the input paired tracks to create realistic transitions.}

\paul{Moreover, instead of using arbitrary deep learning layers to build the generator, based on \emph{differentiable digital signal processing} (DDSP) \cite{ddsp_synthesizer} we propose a novel and light-weight differentiable EQ and fader layers as the core of our generator. This provides a strong inductive bias as the job of the generator now boils down to determining the parameters of the EQ and fader to be applied respectively to the paired tracks.
While DDSP-like components has been applied to audio synthesis \cite{ddsp_synthesizer,ddsp_singing}, audio effect modeling \cite{ddsp_iir,ddsp_blackbox,ddsp_distortion}, and automatic mixing \cite{ddsp_mixing}, 
they have not been used for transition generation, to our best knowledge.}
We refer to our model as DJtransGAN.

\paul{
For model training, we develop a data generation pipeline to prepare input paired tracks from the MTG-Jamendo dataset \cite{mtg_jamendo}, and collect real-world mixes from \emph{livetracklist} \cite{livetracklist}.
\newadd{Examples of the generated mixes and source code can be found at \url{https://paulyuchen.com/djtransgan-icassp2022/}.}} 


\section{Differentiable DJ Mixer}
\label{sec:dj_mixer}
\paul{EQs and faders are two essential components in the DJ-made mixing effects. To make an automatic DJ mixer that is both differentiable and is capable of producing mixing effects, we incorporate DDSP components that resemble EQs and faders in our network. 

We consider audio segments of paired tracks $(x_1,x_2)$, whose lengths have been made identical by zero-padding $x_1$ at the back after $C_\text{out}$ and zero-padding $x_2$ at the front before $C_\text{in}$. We denote their STFT spectrograms as $S_1, S_2 \in \mathbb{R}^{T \times F}$, in which they have $T$ frames and $F$ frequency bins. Two time-frequency masks $M_1(\theta_1), M_2(\theta_2) \in [0.0,1.0]^{T \times F}$, parameterized by $\theta_1$ and $\theta_2$, are used to represent the mixing effects to be applied to $S_1$ and $S_2$. The effects are applied via element-wise product, i.e. $\hat{S_1}=S_1 \odot M_1(\theta_1)$ and $\hat{S_2}=S_2 \odot M_2(\theta_2)$.
The two time-frequency masks $M_1(\theta_1)$ and $M_2(\theta_2)$ represent the DJ mixing effect of fade-out and fade-in, respectively. The goal is to generate $M_1(\theta_1)$ and $M_2(\theta_2)$ such that we can get a proper DJ mix by reverting $\hat{S_1}, \hat{S_2}$ back to the time domain and take their summations as the output $\hat{x}_3$.

}
\begin{figure}[t]
    \centering
    \hspace*{-1cm}
    \includegraphics[width=1.1\linewidth]{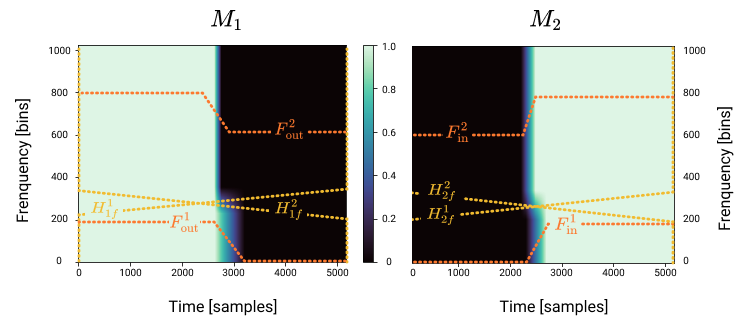}
    \caption{\paul{Differentiable DJ Mixer, illustrating a 2-band EQ.}}
    \label{fig:mixer}
\end{figure}

\subsection{Differentiable Fader}
\label{ssec:fader}
Being inspired by \cite{sigmoid_fader,mobilnet}, we combine two clipped ReLU functions as the basic template of a fading curve, called $PReLU1$. We parameterize $PReLU1$ by its starting time and the slope of its transition region, as formulated below:
\begin{equation}
    \begin{aligned}
        \begin{split}
            & PReLU1(\vec{v_{t}},s_{t},\delta_{t}) = \\ & 
            \min(\max(0,\min(\max(0,\vec{v_{t}}-s_t), 1)*\delta_t), 1) \,,
        \end{split}
    \end{aligned}
    \label{eq:prelu1}
\end{equation}
where $s_{t}$, $\delta_{t}$ and $\vec{v_{t}}$ respectively correspond to the starting time, the slope and a linear increasing sequence ranged from 0 to 1 with $T$ points in the time axis. Accordingly, the fading curves for fade-in and fade-out can be formulated as:

\begin{equation}
    \begin{aligned}
        \begin{split}
            & F_\text{in}(\vec{v_{t}}, \theta_{t})= PReLU1(\vec{v_{t}}, s_{t}, \delta_{t}) \,, \\
             & F_\text{out}(\vec{v_{t}}, \theta_{t})= 1- PReLU1(\vec{v_{t}}, s_{t}, \delta_{t}) \,.
        \end{split}
    \end{aligned}
    \label{eq:fade}
\end{equation}
To ensure that the fading  occurs only inside of the transition region, we add two extra parameters $C_\text{in}$ and $C_\text{out}$, where $0 \leq C_\text{in} \leq C_\text{out} \leq 1$, and constrain that $C_\text{in}\leq s_{t}\leq C_\text{out}$. 
This is similar to the cue button function in an ordinary DJ mixer. 
Likewise, we impose $\delta_{t} \geq \frac{1}{C_\text{out}-s_{t}}$
to prevent the sound volume from reaching the maximum after $C_\text{out}$. 
Note that $F_\text{in}$ and $F_\text{out}$ can be solely determined by $s_{t}$ and $\delta_{t}$; we refer to $F_\text{in}$ and $F_\text{out}$ collectively as $\theta_{t}$ below.

\subsection{Differentiable Equalizer}
\label{ssec:eq}
EQs are used to amplify or attenuate the volume of specific frequency bands in a track. Our network achieves the effect of EQ by decomposing audio into sub-bands with several low-pass filters \cite{subband_separation}, and then using faders to adjust the volume of each sub-band. 

There are several ways to implement a differentiable low-pass filter, such as via FIR filters \cite{ddsp_synthesizer}, IIR filters \cite{ddsp_iir}, or a frequency sampling approach \cite{ddsp_eq,ddsp_distortion}. However, all of them have  limitations. FIR filter requires learning a long impulse response in this case. IIR filter is recurrent, which leads to computational inefficiency. The frequency sampling approach is relatively efficient but requires an inverse FFT every mini-batch during model training. To avoid these issues, we propose to calculate loss in the time-frequency domain to reduce the number of inverse FFT, and we apply fade-out curves along the frequency axis to achieve low-pass filtering. This can be formulated as:
\begin{equation}
    H_{lpf}(\vec{v_{f}}, \theta_{f})= 1 - PReLU1(\vec{v_{f}}, s_{f}, \delta_{f}) \,,
    \label{eq:lowpass}
\end{equation}
where $H_{lpf}$ is the proposed low-pass filter, which can be thought of as ``a fade-out curve in the frequency domain.'' The two parameters $s_{f}$ and $\delta_{f}$ serve similar purposes as the cutoff frequency and the Q factor, which are typical parameters of time-domain filters. $\vec{v_{f}}$ is a linear increasing sequence ranged from 0 to 1 with $F$ points in the frequency axis. 
Moreover, 
to constrain the filtering within a proper frequency band, we introduce $f_{\min}$ and $f_{\max}$ and require  $f_{\min} \leq s_{f}\leq f_{\max}$ and $\delta_{f} \geq \frac{1}{f_{\max}-s_{f}}$.  As $H_{lpf}$ can be solely parameterized by $s_{f}$ and $\delta_{f}$, we refer to them collectively as $\theta_{f}$.

Following the same light, we decompose an input track to in total $k$ sub-bands, and we combine multiple low-pass filters to form a series of fading curves in the frequency domain. We denote the $i$th fading curve as $H_f^{i}$ and the $i$th low-pass filter along with its band limitations as $H_{lpf}^i$, $f_{\min}^i$ and $f_{\max}^i$, with $f_{\max}^{i-1} \leq f_{\min}^i \leq f_{\max}^i$. The $i$th filter implementing the EQ effect, i.e., $H_f^{i}(v_{f}, \theta_{f}^i, \theta_{f}^{i-1})$, can be accordingly formulated as below.  See Figure \ref{fig:mixer} for an illustration.
\begin{equation}
    H_f^{i} = 
    \begin{cases}    
        H_{lpf}^i(\vec{v_{f}},\theta_f^i)\,, & \text{if } i = 1 \,.\\    
        H_{lpf}^i(\vec{v_{f}}, \theta_f^i) - H_{lpf}^{i-1}(\vec{v_{f}}, \theta_f^{i-1})\,, & \text{if } 1 < i < k \,. \\   1 - H_{lpf}^{i-1}(\vec{v_{f}},\theta_f^{i-1})\,, & \text{if } i = k\,. \\
    \end{cases}
     \label{eq:fc}
\end{equation}

\subsection{Differentiable Mixer}
Now we propose a differentiable mixer, it has $k$ sub-bands, and each sub-band has an EQ with its own fader. This means we have two fading-curves for each sub-band, where one works along the time axis and the other along the frequency axis.
We denote $F_\text{in}^{i}$ and $F_\text{out}^{i}$ as the $i$th fade-in and fade-out curves along the time axis, and 
$H_{1f}^{i}$ and $H_{2f}^{i}$ as the $i$th fading curves along the frequency axis. They are parameterized by $\theta_1: \{\theta_{1t}^i,\theta_{1f}^i, \theta_{1f}^{i-1}\}$ and $\theta_2: \{ \theta_{2t}^i,\theta_{2f}^i, \theta_{2f}^{i-1}\}$.

We construct the final fade-in and fade-out effects, namely $M_1(\theta_1)$ and $M_2(\theta_2)$, by summing up the outer products between the corresponding $F^{i}$ and $H_{f}^{i}$ pairs, as shown below.
\begin{equation}
    \begin{split}
        &M_1(\theta_1)=\sum\nolimits_{i=1}^{k} F_\text{out}^i(\vec{v_{t}}, \theta_{1t}^i) \otimes H_{1f}^{i}(\vec{v_{f}}, \theta_{1f}^i, \theta_{1f}^{i-1}) \,,\\
        &M_2(\theta_2)=\sum\nolimits_{i=1}^{k} F_\text{in}^i(\vec{v_{t}}, \theta_{2t}^i) \otimes H_{2f}^{i}(\vec{v_{f}}, \theta_{2f}^i, \theta_{2f}^{i-1}) \,.
    \end{split}
\label{eq:mask}
\end{equation}

In sum, given 
$C_\text{in}$, $C_\text{out}$, $f_{\min}$, and $f_{\max}$, the proposed differentiable mixer has $2 * (k-1)* 4$ trainable parameters, encompassing $\theta_1$ and $\theta_2$. The remaining challenge is to learn proper parameters for the differentiable mixer to generate feasible DJ mixing effects. 

\label{ssec:mask}
\section{Methodology}
\label{sec:methodology}
\begin{figure*}[t]
\centering
    \includegraphics[width=.95\linewidth]{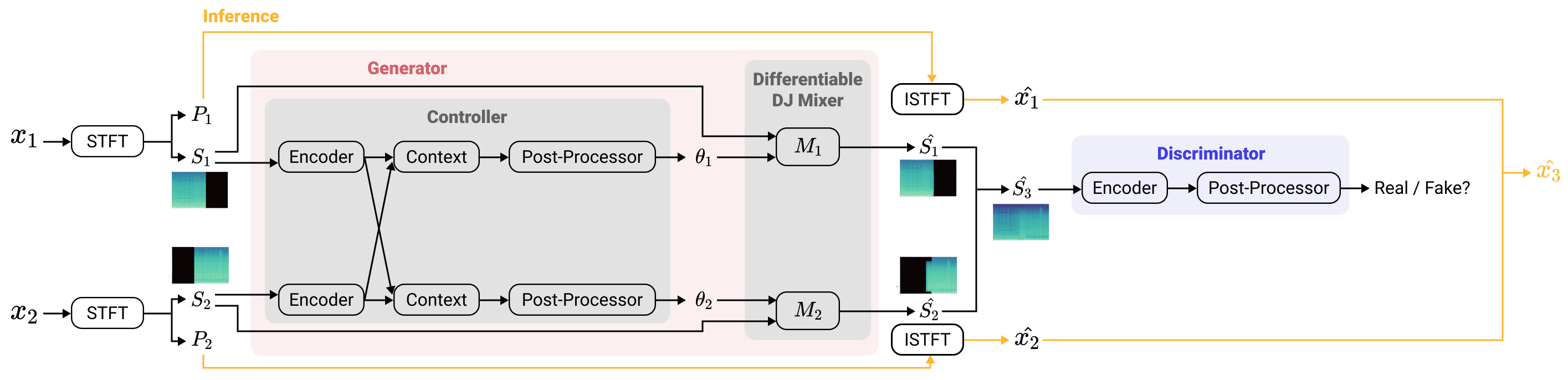}
    \caption{\paul{Schematic plot of the model architecture of the proposed DJtransGAN model for generating DJ transition. }}
\label{fig:model}
\end{figure*}

\subsection{Dataset}
\label{ssec:dataset}

To train our  model, we collect a dataset of real-world mixes, and another dataset of paired tracks with artificially generated cue points.

\textbf{Real DJ transition data.}
\paul{We obtain a 
collection of DJ mixes 
by crawling \emph{livetracklist}, a website hosting DJ mixes \cite{livetracklist}.
We select 284 mixes that use the \emph{mix} tag and based on human-annotated boundaries, and segment the mixes into individual tracks and regard two consecutive tracks as a music section that contains a DJ transition. Sections shorter than 1 minute are discarded. We finally retrieve 7,064 music sections in total.
}

\textbf{Data generation pipeline.}
\paul{
To generate $x_1,x_2$ and the corresponding $C_\text{in},C_\text{out}$, we follow a similar pipeline from \cite{drum&bass}, which uses most of the domain knowledge of DJ practices. We first compute structure boundaries using the constant-Q transform (CQT)-based ``structure feature'' algorithm from MSAF \cite{msaf},  and detect the beat, downbeat, tempo, and musical keys by Madmom \cite{madmom}. Next, we compile a group of segments $x_1$ by extracting those that are between two structure boundaries and with a length greater than 30 bars of music. 
For each $x_1$, we find a suitable $x_2$ to pair with it by firstly selecting 100 candidates that satisfy 1) a bpm difference with $x_1$ no greater than five, 2) a key difference with $x_1$ no greater than two semitones, and 3) originally from a different track. 
Among the 100 candidate, we identify the best fit $x_2$ by the highest mixability score (with $x_1$) as measured by the Music Puzzle Game model \cite{puzzle_game}. Moreover, we pitch-shift and time-stretch $x_2$ to match the pitch and tempo of $x_1$. To fix the transition region to eight bars \cite{highlight_dj,reverse_dj_transition}, we set $C_\text{in}$ and $C_\text{out}$ as the first downbeat of the last eight bars of $x_1$ and the last downbeat of the first eight bars of $x_2$, respectively.
}

\paul{We build a training and a testing corpora respectively with 1,000 tracks and 100 tracks from the songs tagged with \emph{electronic} in the MTG-Jamendo dataset \cite{mtg_jamendo}. In total, we generate 8,318 and 830 paired tracks from each corpus.

}

\textbf{Preprocessing.} 
\paul{
We fix all the inputs to 60 seconds at 44.1 kHz sampling rate. For the real mix dataset, the input is centered to the human-annotated boundary sample. 
For the data from the generation pipeline, we compute the ``cue-mid'' points $C_\text{mid}$ as $(C_\text{in}+C_\text{out})/2$. When needed, music sections are zero-padded symmetrically.

}

\subsection{Model Architecture}
\label{ssec:model}
\paul{
The proposed DJtransGAN model consists of a generator $G$ and discriminator $D$, as shown in Figure \ref{fig:model}. We use $S_1$ and $S_2$ as the input features for both $G$ and $D$ during training. The architecture of $G$ follows the controller network of the Differentiable Mixing Console proposed by Steinmetz \emph{et al.} \cite{ddsp_mixing}. The goal is to learn $\theta_1$ and $\theta_2$, the parameters of the the differentiable DJ mixer.


}

\paul{

Each controller network \emph{encoder} learns a feature map from the respective input $S_1$ or $S_2$. We stack the resulting feature maps and feed them to each context block. The \emph{context} blocks downscale the output channels from two to one and then directly feed them to the post-processors to predict $\theta_1$ and $\theta_2$. $S_1$, $\theta_1$ and $S_2, \theta_2$ are fed to the differentiable DJ mixer to get $\hat{S_1}$ and $\hat{S_2}$. The input to $D$ corresponds to the summation of $\hat{S_1}$ and $\hat{S_2}$.

}

\paul{
The encoder applies a log-scaled Mel-spectrogram layer followed by three residual convolutional blocks. The residual convolutional blocks contain two convolutional layers with a filter size of 3x3 and strides of 2 and 1, respectively. The three residual convolutional blocks have 4, 8, and 16 filters, respectively. Besides, ReLU and batch normalization are applied after all convolutional layers.

}

\paul{
The context block contains a convolutional layer with one filter of size 1x1. The purpose of this block is to provide sufficient cross-information to each post-processor when predicting $\theta_1$ and $\theta_2$. The post-processor contains three MLP layers. Leaky ReLU and batch normalization are applied after all layers except the last MLP layer, where the Sigmoid function is used. The output dimension of each MLP layer is 1,024, 512, and $2 * (k-1)* 4$, where $k$ is the number of bands in the differentiable DJ mixer. The encoder in $D$ is identical to that in the controller network. Similarly, the post-processor in $D$ is the same, although the dimension of the last MLP layer is set to 2. 

}

\begin{figure}[t]
    \centering
    \hspace*{0cm}
    \includegraphics[width=1\linewidth]{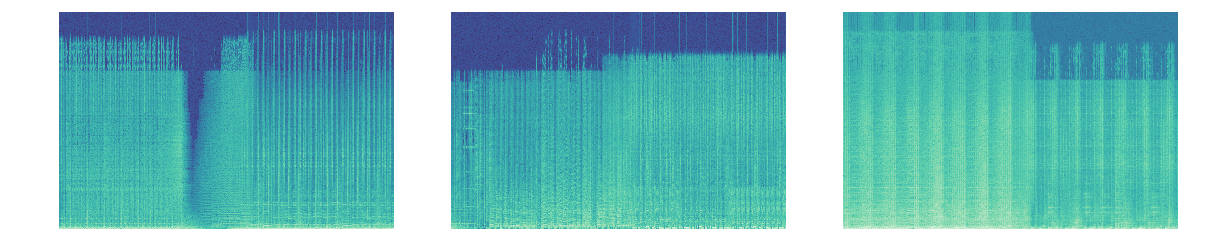}
    \caption{\paul{The STFT spectrograms of some random mixes $(\hat{S_3})$ generated by the DJtransGAN model.}}
    \label{fig:spec}
\end{figure}

\paul{

}

\paul{

}

\subsection{Training and Inference}
\label{ssec:train_inference}

\paul{

In our implementation, we use 128-bin log-scaled Mel-spectrogram computed by using a 2,048-point Hamming window and 512-point hop-size for STFT. We set the differentiable DJ mixers' $k$ as 4, where we set $H_{f}$'s $f_{min}$ as 20, 300, 5,000 Hz and $f_{max}$ as 300, 5,000 and 20,000 Hz, to focus on a low, mid, high-mid and high frequency, respectively. Overall, $G$ is trained to use the differentiable DJ mixer by controlling 24 parameters. We choose a min-max loss \cite{gan} as our training objective and train it with the Adam optimizer for 5,298 steps, batch size of 4, and learning rate 1e$-$5 for both $G$ and $D$.


}

\paul{

As the final step, we apply inverse STFT using the phases $P_1$ and $P_2$ from the inputs $x_1$ and $x_2$, as shown in Figure \ref{fig:model}. We obtain $\hat{x_1}$ and $\hat{x_2}$ as a result and sum them together to get $\hat{x_3}$, i.e., the waveform of the generated transition.
Figure \ref{fig:spec} shows examples of $\hat{S_3}$.


}
\section{Subjective Evaluation}
\label{sec:subjective}

\begin{figure*}[t]
\centering
    \includegraphics[width=1.\linewidth]{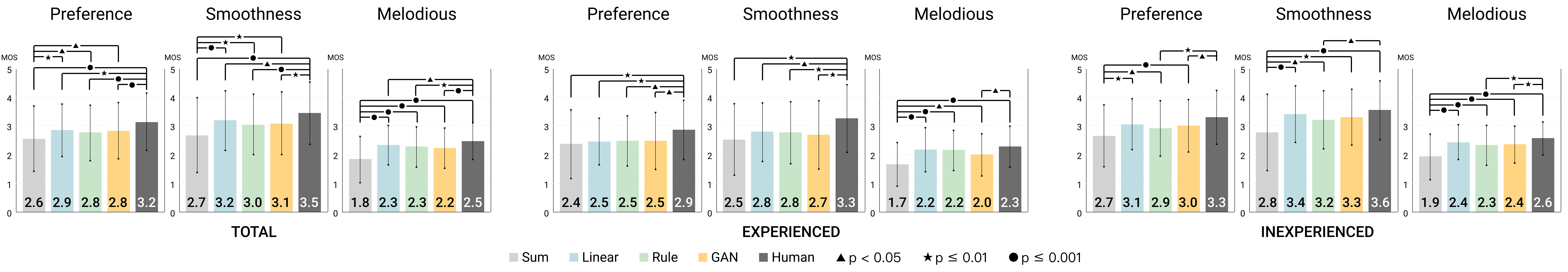}
    \caption{\paul{Result of the subjective evaluation for (left) total people, people (middle) experienced and (right) inexperienced in DJing.}}
\label{fig:subjective}
\end{figure*}

\label{ssec:baseline}
\paul{

We compare our model \emph{GAN} with the following four baselines.
\begin{itemize}[noitemsep,topsep=0pt,parsep=0pt,partopsep=0pt]
\item \emph{Sum} is a summation of $x_1,x_2$ without any effects. 
\item \emph{Linear} applies a linear cross-fading in the transition region which is a common practice in automatic DJ systems \cite{spotify_transition,highlight_dj}. 
\item \emph{Rule} consists of a set of general-purpose decision rules that we devised after consulting with expert DJs. Depending on the presence or absence of vocals, \emph{Rule} applies one of the following four \emph{transitions types}; vocal to vocal (V-V), non-vocal to vocal (NV-V), vocal to non-vocal (V-NV) and non-vocal to non-vocal (NV-NV). For each transition type, a specific fading pattern and EQ preset is applied. 
\item \emph{Human}: lastly, the second author of the paper, who is an ameature DJ, creates the transitions by hand to serve as a possible performance upper bound.  
\end{itemize}
For fair comparison, \emph{Rule} and \emph{Human} use a 4-band EQ.
For all methods and for each pair $x_1,x_2$, the same  $C_\text{in}$ and $C_\text{out}$ are used. 



}

\paul{

We include eight groups of paired tracks from the testing set, two for each transition type (e.g., V-V) mentioned above. We conduct the listening test via an online questionnaire. Participants are directed to a listening page containing the mixes of the five methods for one random group of paired tracks. We first ask them to identify the actual timestamps when they notice the transition, and then to asses each test sample on a five-point Likert scale (1--5) in terms of \emph{Preference} and \emph{Smoothness}, and on a three-point Likert scale (1--3) in terms of \emph{Melodious} (to further test whether the transition contains conflicting sounds).
Participants can also leave comments on each test sample. Finally, we ask them to specify the mix they like the most. We ask this last question mainly to test consistency with the given \emph{Preference} ratings of the individual mixes. 


}
\paul{


}



Figure \ref{fig:subjective} shows the mean opinion score (MOS). 
We discard participants whose answer to the first question is not within the transition region, and those whose answer to the last question shows inconsistency. 
136 out of the 188 participants met the requirements. 
They self-reported their gender (105 male), 
age (17-58), and experience of DJing (46 experienced).
A Wilcoxon signed rank test \cite{wilcoxon} as performed to statistically assess the comparative ability of the five approaches. Overall, the responses indicated an acceptable level of reliability (Cronbach’s $\alpha=$0.799 \cite{cronbach}).

\paul{


}

\paul{
 

Experienced subjects tend to score lower than inexperienced subjects for the transitions. \emph{Human} receives the highest ratings as expected, but there is no statistically significant difference among the methods with respect to {Melodious}. \emph{Linear}, \emph{Rule}, and \emph{GAN} are significantly better than \emph{Sum} for {Melodious}. Inexperienced subjects also rate \emph{Human} the highest, but \emph{Human} is not significant better than \emph{Linear} in Preference and Melodious. 
The difference between \emph{Human} and either \emph{Linear} or \emph{GAN} is also not significant in {Smoothness}. \emph{Sum} receives the lowest ratings as expected.


}

\paul{
We find that some of the experienced subjects tended to give low scores to all transitions. Based on the comments, we infer that the music style of the selected paired tracks is unacceptable to them and therefore greatly influences their ratings. On the other hand, this was not observed in the inexperienced subjects. Overall, there is no significant difference among \emph{GAN}, \emph{Linear} and \emph{Rule}, suggesting that our \emph{GAN} approach can achieve competitive performance compared to the baselines except for \emph{Human}. 



}

\paul{

DJ transitions can be considered an art \cite{spotify_transition,reverse_dj_transition} and therefore a highly subjective task whose result cannot be objectively categorized as correct or incorrect. This is similar to the field of automatic music mixing \cite{eval_mixing}, where an exploration of this issue has led to an analysis of the participants' listening test comments \cite{peer_review}.

Following this idea, we collect in total 150 comments from 58 subjects (24 experienced). 
According to the comments, a large part of the experienced subjects indicate that the paired tracks are not suitable for each other and that the decision of the selected structure and cue points is erroneous, suggesting room for improving the data generation pipeline.
In addition, several comments indicate that these  subjects rate \emph{Linear}, \emph{Rule} and \emph{Human} higher because they can recognize the mixing technique being used. 
In contrast, they are unfamiliar with possible techniques employed by our \emph{GAN}, which may imply that \emph{GAN} creates its own style by integrating multiple DJ styles from real-world data.
Nevertheless, in-depth analysis of the learned mixing style is needed in the future.
Finally, some experienced subjects commend that \emph{GAN} makes good use of the high-pass filter when mixing vocals and background music, especially in the V-NV transition type. The mixes from \emph{GAN} may not be perfect, but they feel more organic than the others (except for the one by \emph{human}). People also criticize \emph{GAN} for making the transition too fast within the mix. These kinds of fast transitions are in particular unpleasant for the V-V type of transition.

}

\section{Conclusion and Future Work}
\label{sec:conclude}
In this paper, we have presented a data-driven and adversarial approach to generate DJ transitions by machine. We have developed a data generation pipeline and proposed a novel differentiable DJ mixer for EQs and loudness faders. Differentiable EQ is achieved in the time-frequency domain by using trainable fade-out curves along the frequency axis, which are based on the frequency response of low-pass filters. Our method is an alternative to differentiable FIR and IIR filters, although we do not have space to present an empirical performance comparison against these alternatives.
We have also conducted a subjective listening test and showed that our model is competitive with baseline models. While not reaching human-level quality, our method shows the feasibility of GANs and differentiable audio effects when performing audio processing tasks. This has potentials to be applied in other tasks such as automatic music mixing and mastering, audio effect modeling, and music synthesis. 

As future work, the quality of the data generation pipeline can be improved, especially the structure segmentation and segment pairing parts. The cue points selection process could also be learned instead of using fixed cue points. \newadd{Global tempo matching can be replaced by individual beat matching.} Furthermore, an analysis of the transitions learned by the model and an exploration of comprehensive evaluation metrics can also be explored.

\bibliographystyle{IEEEbib}
\bibliography{strings,refs}

\end{document}